# Simulation of High Temperature Superconducting Bulks Magnetized by Pulsed Field Magnetization with an Electromagnetic-Thermal Coupled Model

Shengnan Zou, Víctor M. R Zermeño, Francesco Grilli

*Abstract*—High temperature superconducting (HTS) bulks can be magnetized to become powerful trapped field magnets (TFMs), which are promising for high-performance electrical applications. To magnetize such TFMs, pulsed field magnetization (PFM) is supposed to substitute field cooling (FC) to provide *in situ* magnetization. However, the heat generation during PFM, which reduces the trapped field has always been an issue, so numerical simulation of the process is important to provide optimal magnetization strategies. In this paper, HTS bulks magnetized by PFM are simulated with an axisymmetric electromagnetic-thermal coupled model based on *H*-formulation. Influences of important yet difficult-to-characterize parameters of HTS bulks including *n* values in the *E-J* Power Law and $B_0$ in the Kim's Law are investigated. Furthermore, controlled magnetic density distribution coils (CMDCs) which generate a non-uniform field are suggested to further improve the trapped field of HTS bulks compared to split coils proposed previously.

## I. INTRODUCTION

HIGH temperature superconducting (HTS) bulks and stacks of coated conductors can be magnetized to become trapped field magnets (TFMs) by maintaining persistent currents. Such TFMs can provide much higher magnetic flux density than that of ordinary permanent magnets and thus exhibit potential in brushless rotating machines [1]-[4]. To magnetize such TFMs, pulsed-field magnetization (PFM) is a practical solution for a compact and economic machine design; however, PFM will cause considerable heat generation due to fast flux motions and generally achieves lower trapped fields than those acquired by field cooling (FC) or zero field cooling (ZFC) method [5]-[7], especially at lower temperatures. Much work has been reported to simulate such effects in bulks and look for strategies to improve the trapped field of HTS bulks magnetized by PFM, which can be found in a recent review [8]. Simulation of HTS bulks, though more frequently presented in literature than HTS stacks, is more problematic. The HTS bulks are almost impossible to be characterized directly or reliably like HTS tapes. Hypothesis of values of relevant parameters has to be done to carry out simulations.

In this work, we use an axisymmetric electromagnetic-thermal coupled model based on *H*-formulation of Maxwell equations to simulate HTS bulks magnetized by PFM. The influences of crucial parameters to the simulation, including factors *n* in the so-called *E-J* power law and $B_0$ in the Kim's law, are investigated. Besides, PFM using controlled magnetic density distribution coils (CMDCs) is simulated to suggest a further improvement of the trapped field in HTS bulks compared to split coils reported in [9],[10].

## II. MODEL DESCRIPTION

In this work, a 2D finite-element-method (FEM) model describing YBCO bulks based on *H*-formulation [11] of Maxwell equations and heat transfer equation implemented in COMSOL Multiphysics 5.0 [12] is used. The implementation of the *H*-formulation for a 2D axisymmetric model has been described in the authors' group's former work [13]. The electrical resistivity of HTS is given by the so-called *E-J* power law [14],

$$E = E_c \left( \frac{J}{J_c(B,T)} \right)^n, \quad (1)$$

where $E_c$ equals 1e-4 Vm$^{-1}$. For HTS, the parameter *n* usually ranges from 5 (strong flux creep) to 50 (critical state model approximation) [15]. In this work, *n*=8 and 21 will be compared, which were commonly used in simulation of HTS bulks [9,] [16]-[18]. The critical current density $J_c(B, T)$ is magnetic field and temperature dependent, which can be described by the empirical equation named Kim's law [19],

$$J_c(B,T) = J_{c0}(T) \Big/ \left( 1 + \frac{B}{B_0} \right), \quad (2)$$

where $J_0$ equal 0.3 T, 1.3 T and 4 T will be compared as they were typical values used in literature [16], [18], [20]. $J_0(T)$ is the temperature dependent critical current density [8]:

$$J_{c0}(T) = \alpha \left( 1 - \left( \frac{T}{T_c} \right)^2 \right)^{1.5}, \quad (3)$$

where $\alpha$ equal 6.1e8 A/m$^2$ and $T_c$ equal 92 K in this work.

The *H*-formulation is coupled with the heat transfer equation. The heat generation power in the bulk is $E \cdot J$, which increases the temperature; the temperature, in turn, influences the critical current of the HTS bulk as described in Eq. (3). In this model, the temperature dependent heat capacity and

This work has been partly supported by the Helmholtz Association (Young Investigator Group grant VH-NG-617).

S. Zou (e-mail: shengnan.zou@kit.edu), V. Zermeno (e-mail: victor.zermeno@kit.edu), and F. Grilli (e-mail: francesco.grilli@kit.edu), are with Institute for Technical Physics, Karlsruhe Institute of Technology, Hermann-von-Helmholtz-Platz 1, 76344 Eggenstein-Leopoldshafen, Germany.

anisotropic thermal conductivity of the bulk are taken into account referring to the experimental data [21].

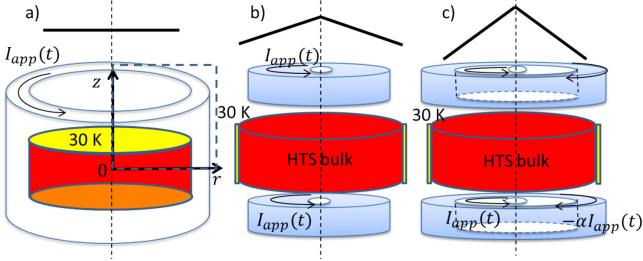

Fig. 1. Schematics of the model of HTS bulks magnetized by PFM with different coils. (a) Solenoid; (b)Vortex coils; (c) Controlled magnetic density distribution coils (CMDCs).

The geometry of the model is shown in Fig. 1. The bulk is a cylinder 50 mm in diameter and 16 mm in thickness. In this model, we only simulate a half of the bulk's cross-section (as marked in Fig. 1a) thanks to the symmetry.

In this work, three different coil configurations are considered. a) The common solenoid, which produces a uniform field. b) The split coils, which were proved to increase the trapped field [9], [10]. Here the coils are 40 mm in diameter, 20 mm thick, and 10 mm apart from the bulk surface. c) The controlled magnetic density distribution coils (CMDCs), which consist of inner turns and outer turns. The coils are 80 mm in diameter, in which the inner turns of coils are 40 mm in diameter. The outer turns carry inversed currents of smaller magnitudes compared to the inner turns; as a result, the generated applied field has a large gradient with the peak value in the center. The idea of CMDCs was discussed for the magnetization of stacks of HTS tapes in the authors' recent work [22]. The ratio between the current density of outer turns and inner turns is defined as $\alpha$ ($0<\alpha<1$), which is 0.15 in this work. The applied field distribution along the bulk radius by three coils configurations in Fig.1 is shown in Fig. 2.

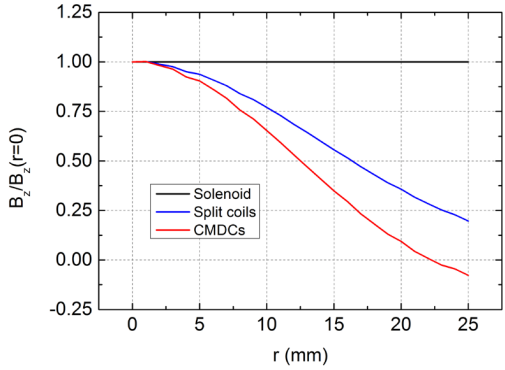

Fig. 2. The normalized applied magnetic field density distribution generated by different coil configurations along the bulk radius (from r=0 to 25 mm) in Fig.1.

The applied field with time is shown in Fig. 3. It ramps for 10 ms and damps for 50 ms. The marked time points (a) to (f) will be used in section III. After the applied pulse, the bulk will relax for 30 s to be cooled to 30 K. The surface or the side of the bulk (dependent on the coils' positions) is attached to a constant 30 K temperature with a thermal resistive (thermal conductivity 0.5 Wm$^{-1}$K$^{-1}$) separation as shown in Fig. 1.

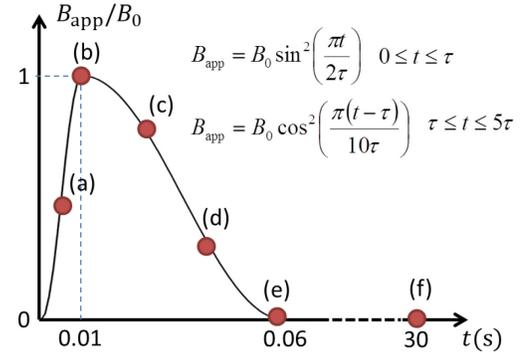

Fig. 3. The time dependence of the applied field. Selected points for analysis are: (a) 0.005 s (b) 0.01 s (c) 0.02 s (d) 0.04 s (e) 0.06 s (f) 30 s

III. RESULTS AND DISCUSSION

In this section, the influence of the $n$ value in the $E$-$J$ power law and $B_0$ in the Kim's law to the simulation of HTS bulks magnetized by PFM will be discussed (subsections *A* and *B*). Such parameters are difficult to be characterized for bulks directly by experiments, so their influences to simulation should be clarified. The simulation of the subsections *A* and *B* is carried out with the geometry in Fig. 1a, where uniform magnetic fields are applied. Then in the subsection *C*, PFM by three different coil configurations in Fig. 1 are compared and discussed to suggest a strategy to further improve the trapped field of HTS bulks magnetized by PFM.

*A. Influence of n values*

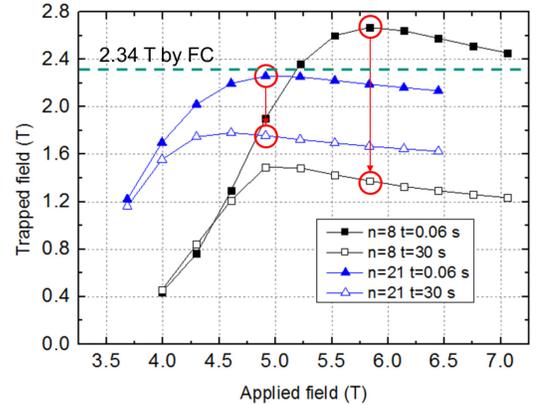

Fig. 4. The trapped field measured 0.8 mm above the centre of the bulk with different magnitudes of applied fields right after the pulse ($t$=0.06 s) and after relaxing ($t$=30 s).

The $B_0$ used for Eq. (2) is fixed as 1.3 T. For this given geometry, the trapped field by FC will be 2.34 T, which is the maximum trapped field reachable theoretically by any magnetization method. The value is acquired by a static model, described in a separate work of the authors' group [23]. The trapped fields with the amplitudes of the applied fields are shown in Fig. 4. The trapped fields are measured 0.8 mm above the bulk's center. The results are shown for $n$=8 or 21, right after the pulse ($t$=0.06 s) or after relaxing ($t$=30 s). For each of these lines, the trapped field first increases and then decreases with the amplitude of the applied field: too small applied fields are not enough to fully magnetize the bulk, while too large applied fields will generate excessive heat and increase the temperature. So there will be an optimal applied field, which is compromised to provide a maximum trapped field.

For $t=0.06$ s, the optimal applied field and the maximum trapped field for $n=8$ (5.8 T) are higher than that for $n=21$ (4.8 T). And the maximum trapped field when $n=8$ is even higher than that acquired by FC (2.34 T). This is because for smaller $n$ values, the flux flow is so strong that the current density is much higher than the critical current, as shown in Fig. 5. For both values, the maximum trapped fields are acquired when there are still inversed currents. The ratio between the maximum trapped field and the optimal applied field is close to 0.5, similar to the case in the ZFC.

The temperature distribution at $t=0.06$ s is shown in Fig. 6. The maximum temperature is found in the periphery of the bulk, because the penetration of currents is from the periphery of the bulk. The maximum temperature is higher for $n=8$ compared to $n=21$.

However, the situation at 0.06 s is transient and it cannot last. What are meaningful in practice are the values at time 30 s, when the bulk is cooled down to 30 K and stabilizes. As shown by the marked points in Fig. 4, the maximum trapped fields at $t=0.06$ s decay substantially during the relaxing. For both $n$ values, the optimal applied fields for $t=30$ s (4.6 T for $n=21$, 4.8 T for $n=8$) are lower than $t=0.06$ s (4.8 T for $n=21$, 5.8 T for $n=8$); because at higher applied fields, the decay is stronger and more initial trapped fields are lost. Finally at $t=30$ s, the ratio between the maximum trapped fields and the optimal applied fields is much smaller than 0.5, which is frequently observed in experiments and explained as a result of flux flow [8].

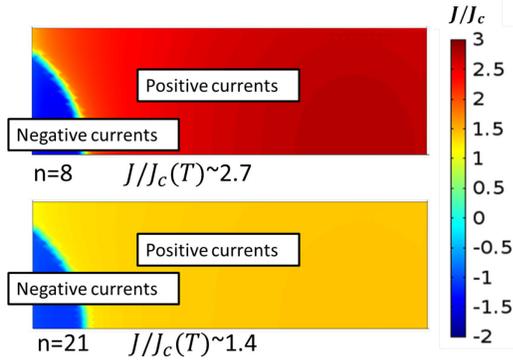

Fig. 5. The normalized current density in the bulk right after the pulse ($t=0.06$ s) when n=8 and n=21, respectively. Only a half of the cross section is plot as marked in Fig. 1a.

In the model, the normalized current density $J/J_c$ can be much larger than 1 during the magnetization. Whether it makes sense physically is not clear. No explicit measurements so far prove that the E-J power law is still valid when $J/J_c$ is much larger than 1. A possible solution is a piecewise function assumption of $n$. The current power law model is a starting approach towards reliable simulation of HTS. The solution mathematically matches Maxwell equations and the non-linear E-J assumption. And basic experimentally observed facts are reproduced by the model: 1. the ratio between the maximum trapped field and the optimal applied field is much less than 0.5, suggesting that much-over-critical currents during magnetization exist and then decay quickly; 2. temperature rising is within a reasonable range. The $n$ value can be regarded as an averaged parameter reflecting the flux flow effects.

### B. Influence of $J_c(B)$

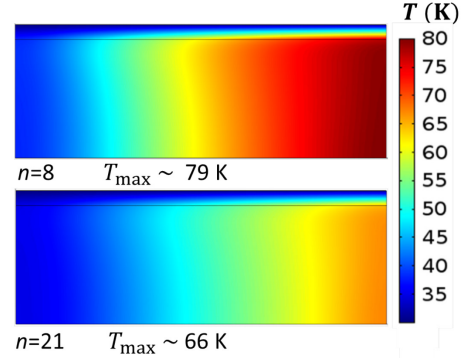

Fig. 6. The temperature distribution right after the pulse ($t=0.06$ s) when n=8 and n=21, respectively. Only a half of the cross section is plot as marked in Fig. 1a.

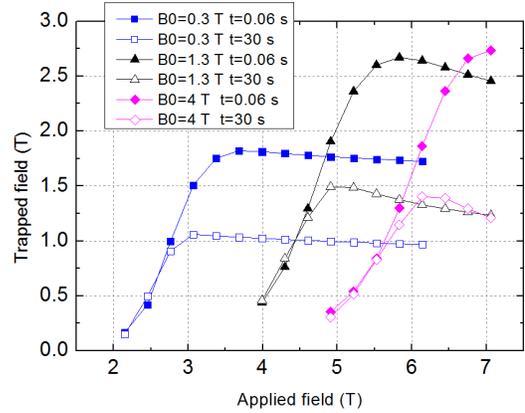

Fig. 7. The trapped field measured 0.8 mm above the centre of the bulk with different magnitudes of applied fields right after the pulse ($t=0.06$ s) and after relaxing ($t=30$ s) for $B_0=0.3$ T, 1.3 T and 4 T, respectively.

In this subsection, the $n$ value is fixed as 8 in Eq. (2). Three different $B_0$ values, 0.3 T, 1.3 T and 4 T in Eq. (3) are used; and the trapped fields acquired by FC are 1.30 T, 2.34 T and 3.37 T, respectively [23]. The trapped fields with the amplitudes of the applied fields are shown in Fig. 7. Similar to the subsection A, the trapped fields decay substantially from $t=0.06$ s to $t=30$ s, and the optimal applied fields for $t=0.06$ s are larger than those at $t=30$ s.

Comparing the results at $t=30$ s for different $B_0$ values, there are several interesting findings. The larger the $B_0$ value, the larger the optimal applied fields; however, the maximum trapped fields first increase and then decrease with $B_0$. This suggests that a sample with better $J_c(B)$ property probably provides yet less trapped fields when it is magnetized by PFM, unlike by FC (here a better $J_c(B)$ property means larger $B_0$ and slower $J_c$ decreasing with magnetic fields). The percentage between the maximum trapped field by PFM and the trapped field by FC is 76%, 63% and 40% for $B_0$ equal 0.3 T, 1.3 T and 4 T, respectively. This means that it is more difficult to make full use of a bulk with better $J_c(B)$ property by PFM. Larger applied fields have to be supplied; however, smaller trapped fields are acquired. In terms of simulation, this result explains how the Kim's law influences the trapped field by PFM.

*C. Coil configuration*

In this subsection, the simulation is carried out for $n=8$ and $B_0=1.3$ T for the three coil configurations in Fig. 1. The trapped fields with the amplitudes of the applied fields are shown in Fig. 8. The maximum trapped fields at $t=30$ s are increased by the split coils and controlled magnetic density distribution coils (CMDCs) compared to the solenoid. Referring to Fig. 2, the larger the spatial gradient of the applied field, the more increase in the maximum trapped field. This was previously explained as a result of heat reduction on the periphery of the bulk by split coils by Fujishiro et al.[9], [10] We also discussed further about influence of the spatial gradient of the applied field on stacks of HTS coated conductors when anisotropic $J_c(B)$ is considered [22]. In this work, anisotropic $J_c(B)$ is not taken into account, an increase of the maximum trapped field by CMDCs is still found.

To compare the penetration processes by the solenoid, split coils and CMDCs, the normalized current distributions for the marked points in Fig. 8 are shown in Fig. 9. For the solenoid, the currents penetrate from the periphery of the bulk; for the split coils, the currents penetrate more from the surface of the bulk; CMDCs push the penetration to the bulk surface even further. As a result, the heat generation is reduced and the temperature distribution is changed, as shown in Fig. 10. The maximum temperature is even more reduced by using CMDCs. The maximum temperatures by CMDCs exist near the surface of the bulk, which is clearly related to the current distribution in Fig. 9e. From time e) to f), the currents redistribute due to flux creep. Finally, the trapped field is increased due to less heat generation and temperature rising. Yet as shown in Fig. 8, the optimal applied for CMDCs are much higher than that of the solenoid, which will add difficulty in generating the applied field. The applicability of such coils needs further research.

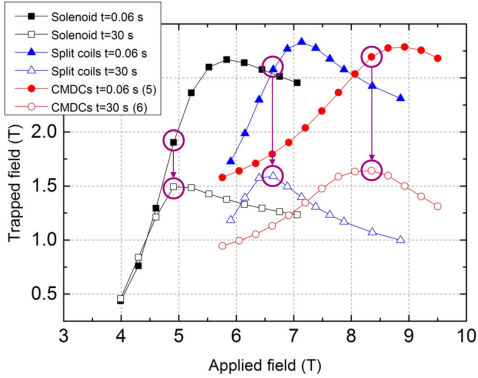

Fig. 8. The trapped field measured 0.8 mm above the centre of the bulk with different magnitudes of applied fields right after the pulse ($t=0.06$ s) and after relaxing ($t=30$ s) for the three coil configurations in Fig. 1.

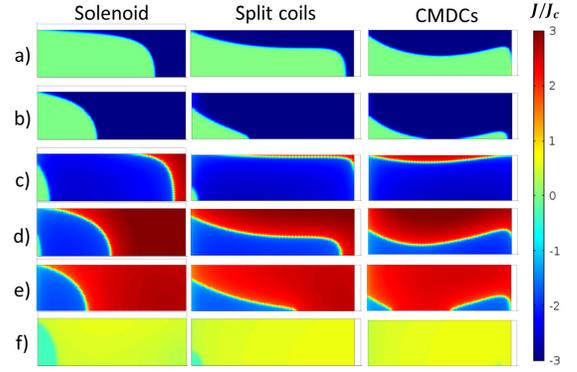

Fig. 9. Penetration processes during PFM of the solenoid, split coils and CMDCs. (a) to (f) correspond to selected time points in Fig. 5. Only a half of the cross section is plot as marked in Fig. 1a.

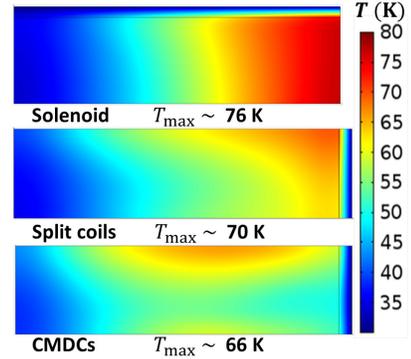

Fig. 10. The temperature distribution right after the pulse ($t=0.06$ s) for the solenoid, split coils and CMDCs, respectively. Only a half of the cross section is plot as marked in Fig. 1a.

## IV. CONCLUSION

In this work, a 2D axisymmetric electromagnetic-thermal coupled model for HTS bulks magnetized by PFM is constructed. The model is based on the *H*-formulation of Maxwell equations and uses a temperature dependent heat capacity and anisotropic thermal conductivity of the HTS bulk. The work first investigates the influences of $n$ in the E-J power law and $B_0$ in the Kim's law on the simulation of HTS bulks magnetized by PFM. These parameters of HTS bulks are difficult to be characterized directly. We found that smaller $n$ values lead to a larger temperature rising and stronger decay after the pulse. And it will result in a smaller ratio between the maximum trapped field and the optimal applied field. For different $B_0$ in the Kim's law, the maximum trapped field first increases and then decreases with $B_0$. The larger the $B_0$ value, the more difficult to make full use of the bulk by PFM. Then the bulk magnetized by the so-called controlled magnetic density distribution coils (CMDCs) is simulated and compared with the common solenoid and previously proposed split coils. The CMDCs produce a highly non-uniform field with the maximum in the center of the bulk. CMDCs reduce the heat generation, so the maximum trapped field by one pulse can be increased by using such coil configurations. Further research needs to be done regarding the feasibility of CMDCs.